\documentclass[aps, amsfonts, amsmath,nofootinbib, showpacs, byrevtex, preprintnumbers, showkeys, preprint, prd]{revtex4-1}   
\usepackage{epsf}
\usepackage{graphicx}

\usepackage{subfigure}
\usepackage{bbm}
\usepackage{showlabels}

\newcommand{\cN}{{\cal N}}

\newcommand{\cD}{{\cal D}}
\newcommand{\cL}{{\cal L}}

\newcommand{\bi}{\bigskip}
\newcommand{\RR}{\mathbbm{R}}
 \newcommand{\no}{\noindent}
\newcommand{\bea}{\begin{eqnarray}}
\newcommand{\eea}{\end{eqnarray}}
\newcommand{\be}{\begin{equation}}
\newcommand{\ee}{\end{equation}}

\newcommand*{\varpm}{\mathbin{\ooalign{\hfil$\pm$\hfil\cr\hfil\raise-.3ex\hbox{$\scriptscriptstyle(\mkern16mu)$}\hfil}}}

\newcommand*{\varmp}{\mathbin{\ooalign{\hfil$\mp$\hfil\cr\hfil$\scriptscriptstyle(\mkern16mu)$\hfil}}}

\newcommand{\dslash}{\partial\hskip -0.5em/}
\newcommand{\lk}{\left(}

\newcommand{\sli}{\sum\limits}

\newcommand{\il}{\int\limits}

\def\R{{\mathbb{R}}}

\newcommand{\ve}{\vec{e}}

\newcommand{\vx}{\vec{x}}
\newcommand{\vz}{\vec{z}}

\newcommand{\vp}{\vec{p}}

\newcommand{\vA}{\vec{A}}

\newcommand{\va}{\vec{a}}

\newcommand{\dk}[1]{\,\,\,\raisebox{-0.4ex}{\large $\bar{}$}\!\!d\,{#1}\,}

\renewcommand{\vec}[1]{\mbox{\boldmath$#1$\unboldmath}}

\renewcommand{\vec}[1]{\mbox{\boldmath$#1$\unboldmath}}
\newcommand*{\ddbar}[1][]{\mathop{\mathrm{d}\mkern-7mu\mathchar'26\mkern-1mu^{#1}}\mkern-4mu}
\begin{document}

\title{Hamiltonian finite-temperature quantum field theory  from its vacuum on partially compactified space}

\author{H.~Reinhardt}
\affiliation{Institut f\"ur Theoretische Physik\\
Auf der Morgenstelle 14\\
D-72076 T\"ubingen\\
Germany}
\date{\today}
%


\begin{abstract}
The partition function of a relativistic invariant quantum field theory is expressed by its vacuum
energy calculated on a spatial manifold with one dimension compactified to a 1-sphere $S^1 (\beta)$, whose
circumference $\beta$ represents the inverse temperature. Explicit expressions for the usual energy density
and pressure in terms of the energy density on the partially compactified spatial manifold $\RR^2 
\times S^1 (\beta)$ are derived. To make the resulting expressions mathematically well-defined a Poisson resummation 
of the Matsubara sums as well as an analytic continuation in the chemical potential are required. The 
new approach to finite-temperature quantum field theories is advantageous in a Hamilton formulation since
it does not require the usual thermal averages with the density operator. Instead, the whole finite-temperature behaviour is encoded in the vacuum wave functional on the spatial manifold $\RR^2 
\times S^1 (\beta)$. We illustrate this approach by calculating the pressure of 
a relativistic Bose and Fermi gas and reproduce the known results obtained from the usual grand canonical ensemble.
As a first non-trivial application we calculate the pressure of Yang-Mills theory as function of the temperature in a quasi-particle
approximation motivated by variational calculations in Coulomb gauge. 
\end{abstract}
\maketitle
Pfad: /paper/unpublished/paper-ham-fin-temp/ham-fin-temp.tex
\bi

\no
\section{Introduction}
\bi

\no
In many branches of modern physics like in the exploration of the early universe or the phase diagram of
hadronic matter the study of a quantum field theory at finite-temperature and chemical potential is 
required. The central quantity of interest is then the grand canonical partition function 
\begin{align}
 \label{1}
 Z (\beta, \mu) &= Tr e^{- \beta (H - \mu N)} 
= \sli_n e^{-  \beta (E_n - \mu N_n)} \, ,
\end{align}
where $\beta$ is the inverse temperature (in units of Boltzmann's constant) and $H$ is the Hamiltonian. Furthermore, $N$ is 
the operator of the number of valence particles and $\mu$ is the corresponding
chemical potential. In the last expression $n$ denotes an exact eigenstate with energy $E_n$ and 
particle number $N_n$. The sum over $n$ includes also the summation over the particle number.
Obviously $Z (\beta, \mu)$ requires the knowledge of all eigenenergies to all possible particle numbers, i.e.
the trace of $\exp (- \beta (H - \mu N))$ in the full Fock space of quantum field theory is required. 
Alternatively, we can represent $Z (\beta, \mu)$ as a functional 
integral over the fields, see eq. (\ref{2}) below. The numerical evaluation of these functional integrals
is the aim of the lattice approach to quantum field theory \cite{R1}. The lattice approach has provided much insight
into the finite-temperature behaviour of quantum field theories and in particular of QCD, where 
extensive calculations have been carried out, for a recent review see e.g. ref. \cite{R2}. 
The lattice approach faces, however, a fundamental problem
when applied to gauge theories at large chemical potentials: the notorious sign 
problem. In SU$(N>2)$ gauge theories the fermion determinant becomes complex for finite chemical potentials, 
which cannot be dealt with in lattice Monte-Carlo calculations. Therefore alternative,
continuum approaches are required for the investigation of QCD at finite baryon density. In the 
continuum, the partition function can be calculated from the functional integral representation in perturbation
theory \cite{R3}, \cite{Kapusta:2006pm}. 
In leading order the partition function is then given in terms of the functional determinants of the inverse propagators
of the fields involved. The perturbative result for the partition function can be extended 
beyond perturbation theory by replacing the bare propagators with  
non-perturbative ones, which are obtained e.g. from a truncated set of Dyson-Schwinger equations \cite{R1DSE} 
or functional renormalization group flow equations \cite{Pawlowski:2014aha}. An alternative non-perturbative continuum approach is the Hamiltonian
approach, which is based on the canonical quantization and requires  the solution of the (functional) Schr\"odiger 
equation \cite{Jackiw:1979ur}. Our experience from quantum mechanics shows that solving the Schr\"odinger equation is 
usually much simpler than calculating the corresponding functional integral, at least when one is only interested in the 
ground state of the theory. Indeed, at zero temperature for QCD in Coulomb gauge the variational Hamiltonian approach developed
in ref.  \cite{Feuchter:2004mk,Reinhardt:2004mm,Epple:2006hv} is much more efficient than the corresponding Dyson-Schwinger approach  \cite{Watson:2006yq}. 
\bi

\no
For finite-temperatures, however, calculating the 
grand canonical partition function via eq. (\ref{1}) seems not very attractive, since, 
in principle, all eigenenergies have to be determined. In the present paper we present a convenient alternative
method to obtain the partition function of a relativistic quantum field theory within
the Hamilton approach where the knowledge of the 
ground state energy (in the presence of a chemical potential) is sufficient, provided the latter is evaluated
on a spatial manifold with one dimension compactified to a circle $S^1 (\beta)$, the circumference of the circle $\beta$
being the inverse temperature. We will present this approach for the generic case of a gauge theory like QCD
with a bosonic vector field and a fermionic Dirac field. The extension to theories with tensor fields will
be straightforward. For a scalar field the approach follows immediately from that of a vector field.
For bosonic fields and in the absence of a chemical potential the present approach to finite-temperature  was 
already used in ref. \cite{Heffner:2015zna} to study Yang-Mills theory. Furthermore, this 
approach also allows the calculation of the Polyakov loop in the Hamiltonian formulation \cite{Reinhardt:2012qe}, \cite{R4A}, 
which at first sight seems impossible due to the use of the Weyl gauge $A_0 = 0$ in the canonical quantization. 
\bi

\no
The organization of the rest of the paper is as follows: In the next section the new Hamiltonian approach to finite-temperature
quantum field theory is developed for a gauge theory with a vector and a Dirac field. In sect. \ref{sectIII} this approach 
is illustrated for free relativistic bosons and fermions.  In sect. \ref{sectIV} the pressure of Yang-Mills theory  is evaluated in 
a schematic quasi-particle model, which is motivated by the results of 
variational calculation in Coulomb gauge \cite{Feuchter:2004mk}, \cite{Epple:2006hv}. Some concluding remarks
are given in sect. \ref{sectV}.

\section{Finite temperature from compactification of a spatial dimension}\label{section2}
\bi

\no
Consider a gauge field theory defined by a classical relativistically invariant Lagrange density $\cL (x; \psi, A; \gamma)$, where
$A^\mu (x)$ and $\psi (x)$, respectively, are the bosonic gauge and the fermionic matter fields. Furthermore,
$\gamma^\nu$ are the Dirac matrices. From the classical Lagrangian one finds
after canonical quantization in the usual way the Hamiltonian $H (\psi, A; \gamma)$. Once the Hamiltonian is known  
the grand canonical partition function (\ref{1}) can, in principle, be calculated.
\bi

\no
From the canonical representation (\ref{1}) of 
the partition function one can derive the following Euclidean functional integral representation, see e.g. 
ref. \cite{Reinhardt:1997rm}
\be
\label{2}
Z (\beta, \mu) = \il_{x^4 - b.c.} \cD (\psi, A) \exp \left[ - S_E [\psi, A] - \mu \il^{\beta/2}_{- \beta/2} d x^4 \int d^3 x \bar{\psi} (x)
\gamma^4 \psi (x) \right] \, ,
\ee
where
\be
\label{3}
S_E [\psi, A] = \il^{\beta/2}_{- \beta/2} d x^4 \int d^3 x \cL_E (x; \psi, A; \gamma)
\ee
is the Euclidean action. The Euclidean Lagrangian 
$\cL_E (x; \psi, A; \gamma)$ follows from the one in  Minkowski space $ \cL (x; \psi, A; \gamma)$ by the analytic continuation
\be
\label{4}
x^4 = i x^0 \, , \quad \quad A^4 = i A^0 \, , \quad \quad \gamma^4 = i \gamma^0 \, ,
\ee
where the Euclidean Dirac matrices satisfy the Clifford algebra
\be
\label{5}
\{ \gamma^\mu, \gamma^\nu \} = - 2 \delta^{\mu \nu} \, .
\ee
In eq. (\ref{2}) the functional integration is performed over temporally (anti-)periodic {(Fermi-) Bose} fields
\be
\label{6}
\psi (\vx, \beta/2) = - \psi (\vx, - \beta/2) \, , \quad \quad A^\mu (\vx, \beta/2) = A^\mu (\vx, - \beta/2) \, .
\ee
These boundary conditions are indicated in eq. (\ref{2}) by the subscript ``$x^4 - b.c.$''. As the derivation of 
eq. (\ref{2}) shows, these boundary conditions are a consequence of the trace in the partition function (\ref{1}).
These boundary conditions are absolutely necessary at finite $\beta$ but become irrelevant   in the zero-temperature limit $\beta \to \infty$.
\bi

\no
Let us now perform the following cyclic change of vectorial variables
\begin{align}
 \label{7}
&z^1 = x^2 \, ,& \quad &z^2 = x^3 \, ,& \quad &z^3 = x^4 \, ,& \quad &z^4 = x^1 \nonumber\\
&C^1 (z) = A^2 (x) \, ,& \quad &C^2 (z) = A^3 (x) \, ,& \quad &C^3 (z) = A^4 (x) \, ,& \quad &C^4 (z) = A^1 (x)  \nonumber\\
&\Gamma^1 = \gamma^2 \, ,& \quad &\Gamma^2 = \gamma^3 \, ,& \quad &\Gamma^3 = \gamma^4 \, ,& \quad &\Gamma^4 = \gamma^1 
 \end{align}
and also, change the fermion variable, by the identification
\be
\label{296-*}
\chi (z) = \psi (x) \, .
\ee
The new Dirac matrices $\Gamma^\mu$ satisfy the same Clifford algebra  (\ref{5}) as the 
old ones, $\gamma^\mu$, and thus the same matrix representation can be used for the $\Gamma^\mu$ as for the $\gamma^\mu$. 
The change of variables (\ref{7}), (\ref{296-*}) can be accomplished by a particular $O (4)$ rotation. Therefore, by the 
$O (4)$ invariance of the Euclidean Lagrangian we have\footnote{Strictly speaking, this identity requires that the same representation is chosen 
for the new Dirac matrices $\Gamma^\mu$ as for the old ones
$\gamma^\mu$. However, since the physical quantities are independent of the specific representation used we can employ any representation for 
the $\Gamma^\mu$, which is convenient. }
\be
\label{8}
\cL_E (x; \psi, A; \gamma) = \cL_E (z; \chi, C; \Gamma) \, .
\ee
After the change of variables (\ref{7}) the functional integral for the partition function (\ref{2}) becomes 
\be
\label{9}
Z (\beta, \mu) = \il_{z^3 - bc} \cD (\chi, C) \exp \Big[ - S_E [\chi, C] - \mu \int d z^4 \il_\beta d^3 z \bar{\chi} (z) \Gamma^3 \chi (z)
\Big]  \, ,
\ee
where we have defined the integration measure over the partially compactified spatial manifold $\RR^2 \times S^1 (\beta)$ 
\be
\label{272-5}
\il_\beta d^3 z := \int d^2 z_\perp \il^{\beta/2}_{- \beta/2} d z^3 
\ee
with $\vz_\perp = (z^1, z^2)$ denoting 
the components of the three-vector orthogonal to the compactified dimension $z^3$. Furthermore, the Euclidean action is now given by
\be
\label{11}
S_E [\chi, C] = \int d z^4 \il_\beta d^3 z \cL_E (z; \chi, C; \Gamma) 
\ee
and the functional integration runs over fields satisfying (anti-)periodic boundary condition in the $z^3$ direction
\be
\label{12}
\chi (\vz_\perp, \beta  / 2, z^4) = - \chi (\vz_\perp, - \beta/2, z^4), \quad \quad C^\mu (\vz_\perp, \beta/2, z^4) = C^\mu (\vz_\perp, - \beta/2, z^4) \, ,
\ee
which is indicated in eq. (\ref{9}) by the subscript ``$z^3 - bc$''. 
\bi

\no
We can now interpret $z^4$ as Euclidean time and $\vz = (\vz_\perp, z^3) = (z^1, z^2, z^3)$ as the spatial
coordinates and reverse the derivation
which leads from the canonical representation (\ref{1}) to the path integral representation (\ref{2}). Then we arrive at the following canonical
representation of the grand canonical partition function
\be
\label{13}
Z (\beta, \mu) = \lim\limits_{l \to \infty} Tr e^{- l \tilde{H}   (\chi, C; \Gamma; \beta, \mu)} \, 
\ee
where $l \to \infty$ is the length of the uncompactified spatial dimensions and
\be
\label{14}
\tilde{H} (\chi, C; \Gamma; \beta, \mu) = H (\chi, C; \Gamma; \beta) + i \mu \il_\beta d^3 z \chi^\dagger (z) \alpha_3 \chi (z) \, .
\ee
Here $H (\chi, C; \Gamma; \beta)$ is the Hamiltonian, which arises 
after the analytic continuation 
\be
\label{15}
z^4 = i z^0 \, , \quad \quad C^4 = i C^0 \, , \quad \quad \Gamma^4 = i \Gamma^0 
\ee
of the Euclidean Lagrangian $\cL_E (z; \chi, C; \Gamma)$ to Minkowski space and by subsequent  
canonical quantization in ``Weyl gauge'' $C^0 (z) = 0$ (considering 
$C^{i = 1, 2, 3} (z)$ as the ``coordinates'' of the gauge field).  Furthermore, we have defined the Dirac matrix 
\be
\label{16}
\alpha_3 = \Gamma^0 \Gamma^3 \, .
\ee
The resulting Hamiltonian $H (\chi, C; \Gamma; \beta)$ is formally the same as the original Hamiltonian $H (\psi, A; \gamma)$ except that the 
former is defined on the spatial manifold $\RR^2 \times S^1 (\beta)$ and its fields satisfy the (anti-)periodic spatial boundary conditions 
(\ref{12}). 
\bi

\no
The representation (\ref{13}) allows one to calculate the partition function in an efficient 
way within the Hamiltonian approach: Due to the limit $l \to \infty$, arising from the infinite extent of the  (original) spatial dimensions,
the calculation of the function (\ref{13}) reduces 
to finding the vacuum energy $\tilde{E}_0 (\beta, \mu)$ of the quantum field theory defined by the Hamiltonian (\ref{14})
\be
\label{17}
Z (\beta, \mu) = \lim\limits_{l \to \infty} e^{- l \tilde{E}_0 (\beta, \mu)} \, .
\ee
This requires to solve the Schr\"odinger equation 
\be
\label{18}
\tilde{H} (\beta, \mu) \psi (\chi, C) =  \tilde{E}_0 (\beta, \mu) \psi (\chi, C) 
\ee
for the vacuum wave functional on the spatial manifold $\RR^2 \times S^1 (\beta)$ for given $\beta$ and $\mu$.  In this way the whole 
finite-temperature behaviour of the quantum field theory is encoded in the vacuum sector on the spatial manifold $\RR^2 \times S^1 (\beta)$.
The upshot of the above consideration is that finite-temperature quantum field theory can be described in the Hamiltonian approach by 
compactifying one spatial dimension and solving  the corresponding Schr\"odinger equation for the vacuum sector. Let us stress that 
the equivalence between eq. (\ref{1}) and eqs. (\ref{13}), (\ref{17}) holds for any relativistically 
invariant theory. In the derivation of eq. (\ref{13}) 
we have used the $O (4)$-invariance of the Euclidean Lagrangian. Consequently, the present approach to 
finite-temperatures cannot be applied to non-relativistic field theories or manybody systems. 
\bi

\no
In terms of the original partition function (\ref{1}) the pressure $p$ and energy density $\varepsilon$ are given by 
\begin{align}
 \label{19}
 p &= \ln Z (\beta, \mu) / \beta V \nonumber\\
 \varepsilon &= \langle H \rangle / V = \frac{1}{V} 
 \lk - \frac{\partial \ln Z}{\partial \beta} + \frac{\mu}{\beta} \frac{\partial \ln Z}{\partial \mu} \right)  \, ,
 \end{align}
where $V = l^3 \, , \, l \to \infty$ is the volume of ordinary 3-space. Inserting here for  $Z$ the representation (\ref{17})
one finds 
\be
\label{20}
p = - e \, , \quad \quad \varepsilon = \frac{\partial}{\partial \beta} (\beta e) - \mu \frac{\partial e}{\partial \mu} \, ,
\ee
where $e$ denotes the vacuum energy density on $\RR^2 \times S^1 (\beta)$  defined by 
\be
\label{21}
\tilde{E}_0 (\beta, \mu) = l^2 \beta e \, .
\ee
To distinguish this quantity from the true (physical) energy density $\varepsilon$ in the following we will refer to ${e}$ as 
{\em pseudo-energy density}.
Analogously one finds from
\be
\label{369-f7}
\langle  N \rangle = \frac{1}{\beta} \frac{\partial \ln Z}{\partial \mu}
\ee
and eqs. (\ref{17}), (\ref{21}) for the particle density $\rho = \langle N \rangle / V$
\be
\label{374-7}
\rho = - \frac{\partial e}{\partial \mu} \, ,
\ee
which together with eq. (\ref{20}) yields the known result
\be
\label{379-7a}
\rho = \frac{\partial p}{\partial \mu} \, .
\ee
The above presented approach to finite-temperature quantum field theory is completely equivalent to the standard grand canonical ensemble as 
long as no approximation is introduced that breaks the relativistic invariance. 
It is, however, advantageous in non-perturbative studies, since
it requires only the calculation of the ground state energy density on the spatial manifold $\RR^2 \times S^1 (\beta)$ 
but avoids the introduction
of the density operator $\exp (- \beta ({H} - \mu N))$ of the grand canonical ensemble. The latter quantity is difficult to 
handle in a continuum approach when strong interactions are present. 
\bi

\no
\section{Illustration for free fields}\label{sectIII}
\bi

\no
Below we illustrate the approach to finite-temperature quantum field theory presented in the previous section for free 
(relativistic) field theories.
\bi

\no
In the usual functional integral approach the partition function of the grand canonical
ensemble of a free field theory is obtained in terms of the functional determinant 
of the inverse Euclidean propagator \cite{Kapusta:2006pm}. For example, for a massive complex scalar Bose field one finds
\be
\label{375-22}
Z_B (\beta) = Det^{- 1} (- \partial_\mu \partial^\mu - m^2) \, ,
\ee
while for a massive Dirac fermion field
\be
\label{380-23}
Z_F (\beta) = Det (- i \dslash - m)
\ee
is obtained. In both cases $m$ denotes the mass. 
Here the eigenfunctions of the inverse propagators have to satisfy the \mbox{(anti-)}periodic boundary conditions for (fermions) bosons as 
a consequence of the temporal (anti-)periodic boundary condition (\ref{6}) to the fields. The functional determinants are UV-divergent
and a few 
mathematical manipulations (like partial integration and dropping infinite temperature-independent constants) are required in order to obtain
from the partition function well-defined expressions for the thermodynamic quantities. Therefore 
in the alternative approach of the previous section we should also expect that
some mathematical manipulations are required to obtain well-defined expressions. 

With eq. (\ref{19}) one finds from eqs. (\ref{375-22}, \ref{380-23}) for
the pressure (see e.g. ref. \cite{Kapusta:2006pm})
\be
\label{388-24}
p = \cN \frac{1}{3} \int d^3 p \frac{p^2}{\omega (p)}  \lk n_+ (p) + n_- (p) \right) \, ,
\ee
where $\cN$ is a numerical factor accounting for the  number 
of degenerate degrees of freedom\footnote{For a scalar field $\cN = 1$ while $\cN = 2$ for Dirac fermions due to the two 
degenerate spin states.} and 
\be
\label{393-25}
n_\pm (p) = \left[ e^{\beta (\omega (p) \mp \mu )} \varmp 1 \right]^{- 1} \, ,
\ee
are the finite-temperature (Fermi) Bose occupation numbers of the particle and anti-particles, respectively. Here, 
\be
\label{459-29}
\omega (p) 
= \sqrt{\vp^2 + m^2}
\ee
is the relativistic single particle energy for both bosons and fermions. 
The same result, eq. (\ref{388-24}), is also found in the usual Hamiltonian approach by calculating the grand canonical partition function 
(\ref{1}) in the corresponding Fock space with $H$ given by the free particle Hamiltonian. 
\bi

\no
For a real field the chemical potential vanishes such that $n_+ (p) = n_- (p) := n (p)$. In addition, a factor of $1/2$
arises due to the fact that the partition function is then given by only the (inverse) 
square root of the functional determinant of the inverse propagator, i.e. for a real scalar field, one has
\be
\label{481-f9}
Z_B (\beta) = Det^{- 1/2} \lk - \partial_\mu \partial^\mu + m^2 \right) 
\ee
such that
\be
\label{486-f9-1}
p = \cN \frac{1}{3} \int \ddbar^3 p \frac{p^2}{\omega (p)} n (p) \, 
\ee
with $\cN = 1$.
This expression applies also for massless gauge bosons where $\cN = 2$ accounts for the two polarization degrees of freedom. 
\bi

\no
Let us now calculate the pressure in 
the alternative approach developed in sect. \ref{section2}. In this approach the central quantity of interest is the vacuum
energy density on $\RR^2 \times S^1 (\beta)$. For free field theories this quantity is of the form
\be
\label{401-25}
e (\beta) = (-)^{n_F} \cN \frac{2^{n_F}}{2}  \il_\beta \dk  {}^3 p \, \Omega (\vp) \, ,
\ee
where 
we have introduced the fermion number
\be
\label{500-10}
n_F = \left\{ \begin{array}{lll}
0 & , & \mbox{bosons} \\
1 & , & \mbox{fermions}
              \end{array} \right.
              \ee
 in order to treat Bose and Fermi systems simulteneously.
 Furthermore, $\Omega ({\vp})$ is a generalized single particle energy.\footnote{For Fermi system 
the vacuum is given by the filled negative energy states of the Dirac sea. With the sign convention adopted in eq. (\ref{401-25}) 
$\Omega (\vp)$ is positive also for fermions. The additional factor of $1/2$ in the Bose case $(n_F = 0)$ arises from the 
ground state energy of an independent oscillator mode.}  (For a massive 
free particle $\Omega (\vp)$ is given by $\omega (p)$ (\ref{459-29}).) It 
is a function of the 3-momentum $\vp$, which on $\R^2  \times  S^1 (\beta)$ is given by 
\be
\label{406-26}
\vp = \vp_\perp + p_n \ve_3 \, .
\ee
Here $\vp_\perp$ denotes the component of the momentum in the two non-compactified 
spatial dimensions, while 
\be
\label{412-27}
p_n = \omega_n + n_F \frac{\pi}{\beta} \, , \quad \quad \omega_n = \frac{2 \pi n}{\beta} \,
\ee
denotes the Matsubara frequencies for Bose $(n_F = 0)$ and Fermi $(n_F = 1)$ 
systems. Furthermore, we have defined the integration measure of the corresponding
momentum space 
\be
\label{422-28}
\il_\beta \dk {}^3 p = \int \dk {}^2 p_\perp \frac{1}{\beta} \sli_n \, , \quad \quad \dk {}^2 p_\perp = \frac{d^2 p_\perp}{(2 \pi)^2} \, .
\ee
According to eq. (\ref{20}) the expression (\ref{401-25}) should give for $\Omega (p) = \omega (p)$ the negative of the pressure 
given in eq. (\ref{388-24}), which is  
not immediate obvious. Contrary to eq. (\ref{388-24})  
the expression (\ref{401-25}) is UV-divergent. This is not surprising since $e$ contains the (infinite) zero 
temperature vacuum energy density, which has to be eliminated from the thermodynamic quantities. To extract the zero-temperature part of $e (\beta)$
(\ref{401-25})
it is convenient to Poisson resum the sum over the Matsubara frequencies using 
\be
\label{429-29}
\frac{1}{\beta} \sli^\infty_{n = - \infty} f (\omega_n) = \frac{1}{2 \pi} \il^\infty_{- \infty} d z f (z) \sli^\infty_{k = - \infty}
e^{i k \beta z} \, .
\ee
Putting $z = p_3$ for bosons and $z + \frac{\pi}{\beta} = p_3$ for fermions 
we obtain for any function of the norm of the 3-momentum $f (\vp) \equiv f \lk \sqrt{\vp^2_\perp + p^2_n} \right)$ the relation
\be
\label{495-10}
\il_\beta \dk {}^3 p f (\vp) = \int \dk {}^3 p f (\vp) \sli^\infty_{k = - \infty} (-1)^{n_F k} e^{i k \beta p_3} \, .
\ee
              Note on the l.h.s. the integration includes the Matsubara sum (\ref{422-28}), while on the r.h.s.
              the integration is over the usual three dimensional (flat) momentum space with $d^3 p = d^3 p / (2 \pi)^3 \, , d^3 p = d  p_1 d p_2
              d p_3$. 
              The $k = 0$ term in the sum (\ref{495-10})  represents just the zero 
              temperature part. In a quantum field theory this part is   usually divergent and has to 
              be eliminated to find the temperature-dependent
              part. Even when the $k = 0$ term is excluded for the 
              functions $f (\vp)$ of interest 
              the momentum integral on the r.h.s. of eq. (\ref{495-10}) is usually UV-divergent. To make these integrals well defined 
              we use the proper-time representation for a power
              of a  quantity\footnote{
From the integral representation of the incomplete $\Gamma$-function
\be
\label{492-32}
\Gamma (\nu, x) = \il^\infty_x d s s^{\nu - 1} e^{- s}
\ee
 the representation (\ref{515-10-2}) follows after the substitution $s = \tau A$.
} 
 $A$            
\be
\label{515-10-2}
A^{\nu} = \frac{1}{\Gamma \lk - \nu, x \right)} \il_{x/A} d \tau \tau^{- 1 - \nu} e^{- \tau A} \, ,
\ee
which is valid for any $x > 0$. However, this representation is not yet useful since the quantity of interest $A$
appears also in the lower integration bound.
As we will see below, after the momentum integrals have 
been carried out the limit $x \to 0$ can be taken in the proper-time integral, which removes $A$ from the integration bound. 
However, the incomplete $\Gamma$-function
$\Gamma (\nu, x)$ is divergent for $\nu < 0$ and $x \to 0$.
The regularization can be carried out as usual by replacing the incomplete $\Gamma$-function $\Gamma 
\lk - \nu, x \right)$ in the limit $x \to 0$ by its complete counterpart $\Gamma \lk - \nu \right)$ provided $\Gamma (- \nu)$ does exist.
(This in fact is an analytic continuation, which discards an infinite contribution.)
Then we obtain from (\ref{515-10-2}) the representation
\be
\label{482-9}
A^{\nu} = \frac{1}{\Gamma \lk - \nu \right)} \lim\limits_{\epsilon \to 0} \il^\infty_\epsilon 
d \tau  \tau^{- 1  - \nu} e^{- \tau A} \, .
\ee

For the energy density (\ref{401-25}) with a general dispersion relation $\Omega (p)$ we find by using eq.  (\ref{495-10}) 
\be
\label{436-30}
e (\beta)   = (-)^{n_F}  \cN \frac{2^{n_F}}{2} \int \dk {}^3 p \Omega (\vp) \sli^\infty_{k = - \infty} (-)^{k n_F} e^{i k \beta p_3} \, . 
         \ee
 As already noticed before, the $k = 0$ term is just the zero temperature part of the vacuum energy density, which, indeed, is a 
              (infinite) temperature-independent 
              and thus irrelevant constant, which has to be omitted from thermodynamical quantities.
       The remaining temperature-dependent part is still not well-defined. At least the integral over the transverse momenta $\int \dk {}^2 p_\perp$ 
is still UV-divergent. To make this integral well-defined we use the proper-time representation (\ref{482-9}) with
$A = \Omega^2$ and $\nu = 1/2$.
Then we find from eq. (\ref{436-30}) for the finite-temperature part of the vacuum energy on 
$\RR^2 \times S^1 (\beta)$
\begin{align}
\label{513-34}
e (\beta)  &= (-)^{n_F}  \cN  \frac{2^{n_F}}{2}  \frac{2}{\Gamma \lk - \frac{1}{2} \right)} \il^\infty_0 d \tau 
\tau^{- 3/2} \int \dk {}^3 p e^{- \tau \Omega^2 (\vp)}  \sli^\infty_{k = 1} (-)^{k n_F} \cos (k \beta p_3)
\end{align}
For $\beta \to \infty$ the integrand becomes a rapidly oscillating function and $e (\beta)$ vanishes, as one expects for the finite-temperature part. 
To work out the expression (\ref{513-34}) we need the explicit form of the single particle energies $\Omega ({\vp})$.
We will consider bosons and fermions separately. For simplicity we will put the chemical potential to zero in the Bose case, since we are 
mainly interested in gauge bosons, where the chemical potential vanishes. For fermions we will include a non-zero chemical potential. 
\bi

\no
\subsection{Bosons}
\bi

\no
For massive relativistic bosons and in the absence of a chemical potential the quantity $\Omega (\vp)$ in 
eq. (\ref{401-25}) is given by the single particle energy (\ref{459-29}) 
\be
\label{534-35}
\Omega (\vp) = \sqrt{\vp^2 + m^2} \, .
\ee
With this form of $\Omega ({\vp})$ the momentum integrals in eq. (\ref{513-34}) can be explicitly carried out. Using $\Gamma \lk - \frac{1}{2} \right) = - 2 \sqrt{\pi}$ 
one finds 
\be
\label{540-36}
e  (\beta) = \frac{\cN}{8 \pi^2} \sli^\infty_{n = 1} \lk \frac{2 m}{n \beta} \right)^2 K_{- 2} (n \beta m) \, ,
\ee
where
\be
\label{545-37}
K_\nu (z) = \frac{1}{2} \lk \frac{1}{2} z \right)^\nu \il^\infty_0 d t t^{- \nu - 1} e^{- t - \frac{z^2}{4 t}}
\ee
is the modified Bessel function, which satisfies the relation $K_{- \nu} (z) = K_\nu (z)$. 
\bi

\no
For massless bosons $m = 0$ the expression (\ref{540-36}) can be worked out analytically. Using the asymptotic form
of the modified Bessel function \cite{R7}
\be
\label{558-38}
K_\nu (z) = \frac{1}{2} \Gamma (\nu) \lk \frac{1}{2} z \right)^{- \nu} \, , \quad \quad z \to 0 \, , \quad \quad \nu > 0 \, 
\ee
we find from eq. (\ref{540-36}) for $m = 0$ 
\be
\label{560-39}
e (\beta)  = -  \frac{\cN}{\pi^2} \zeta (4) T^4 \, ,
\ee
where 
\be
\label{632-13}
\zeta (x) = \sli^\infty_{n = 1} \frac{1}{n^x}
\ee
is the Riemann $\zeta$-function.
With $\zeta (4) = \pi^4 / 90$ we obtain from (\ref{20}) for the pressure of massless bosons 
\be
\label{570-41}
p = \cN \frac{\pi^2}{90} T^4 \, , \quad \quad T = 1 / \beta \, ,
\ee
which is the Stefan-Boltzmann law. This is the correct result, which follows also from the usual grand canonical ensemble,
eq. (\ref{388-24}). Note for photons we have $\cN = 2$ due to the two polarization
degrees of freedom.
\bi

\begin{figure}
 \includegraphics[width=.8\textwidth]{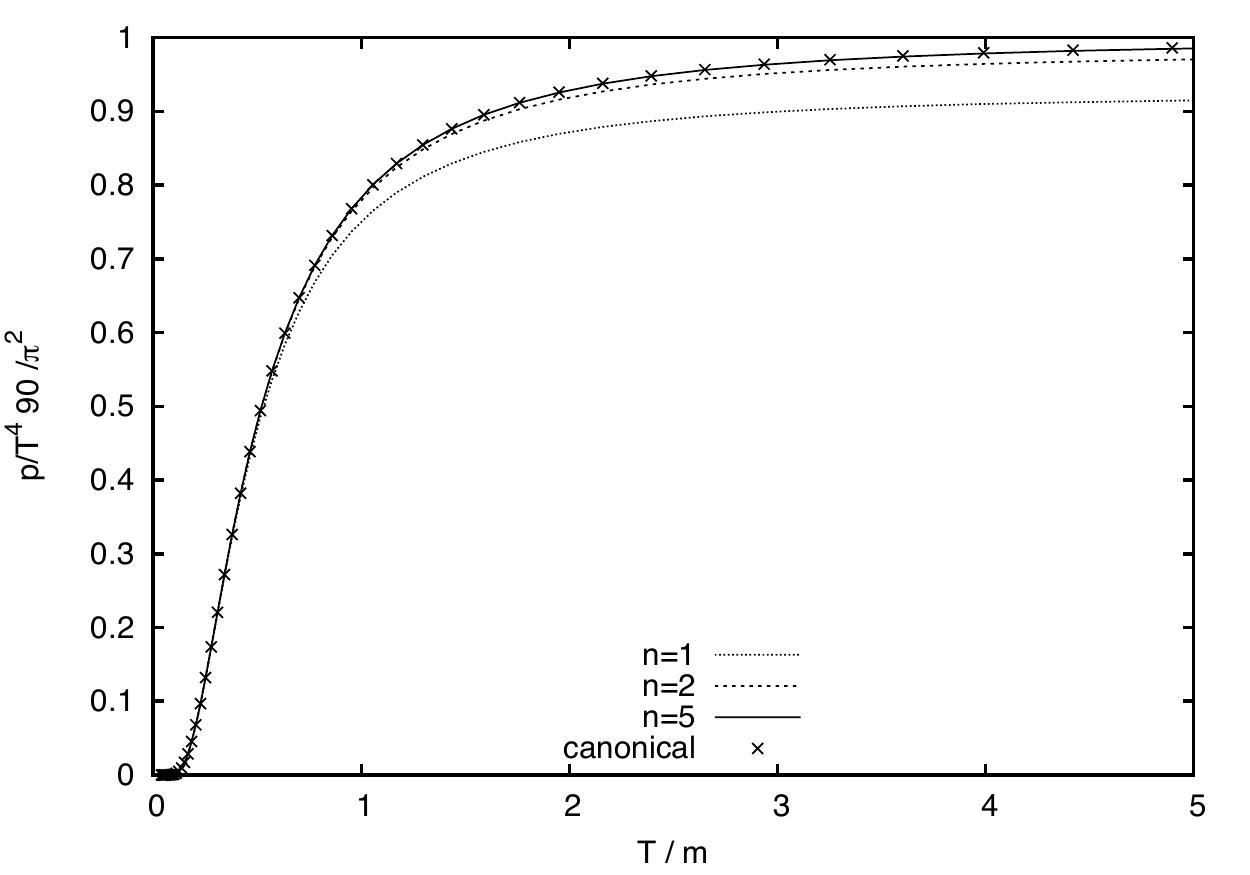}
  \caption{The pressure of a gas of massive bosons as function of the temperature calculated from eq. (\ref{540-36}) 
 with $\cN = 1$ 
 by including only the first $n = 1, 2$ and $5$  terms. The crosses give the result of the grand canonical ensemble (\ref{388-24}).}\label{fig1}
\end{figure}

\no
For massive bosons $m \neq 0$ the remaining expression (\ref{540-36}) has to be calculated numerically. Also the corresponding
expression  (\ref{388-24}) for the pressure in the usual grand canonical ensemble cannot be calculated analytically
in that case. Fig. \ref{fig1} shows 
the pressure calculated numerically by means of eq. (\ref{540-36}) and compares this result to that of the grand canonical 
ensemble (\ref{388-24}). The agreement of both results is, of course, expected  in view of our above given derivation. What is remarkable, 
however, is that only the first few terms of the sum in eq. (\ref{540-36}) are required to reproduce the full result with high accuracy.

\bi

\no
\subsection{Fermions}
\bi

\no
In the present approach of introducing the temperature through the compactification of the 3-axis the Dirac Hamiltonian of 
massive fermions in the presence of a (real) chemical potential reads, see eq. (\ref{14})
\be
\label{592-42}
h = \vec{\alpha} \cdot \vp + \gamma^0 m + i \mu \alpha_3 \, , \quad \quad \alpha_k = \gamma^0 \gamma^k \, .
\ee
Its eigenvalues are given by $\pm \Omega (\vp)$ where
\be
\label{597-43}
\Omega (\vp) = \sqrt{m^2 + \vp^2_\perp + (p_n + i \mu)^2} \, 
\ee
with $p_n = \omega_n + \pi / \beta$ being the fermionic Matsubara frequency, eq.~(\ref{412-27}). 
Inserting this expression for $\Omega (\vp)$ into eq. (\ref{513-34}) and carrying out the momentum integral analogously to the 
bosonic case,
one arrives at the following result 
\be
\label{604-44}
e (\beta) = \frac{\cN}{4 \pi^2} \sli^\infty_{n = 1} (-)^n \cos (i n \beta \mu) \lk \frac{2m}{n \beta} \right)^2 K_{- 2} (n \beta m) \, ,
\ee
where we have $\cN = 2$ due to the two spin degrees of freedom for Dirac fermions. For massive fermions this expression has to 
be calculated numerically, while
for massless fermions we can use the asymptotic form of the modified Bessel function (\ref{558-38}) and obtain 
\be
\label{609-45}
e (\beta) = \frac{2 \cN}{\pi^2 \beta^4} \sli^\infty_{n = 1} (-)^n \frac{\cos (i n \beta \mu)}{n^4} \, .
\ee
Obviously, this sum does not converge for real $\mu$ and $\beta$. To make 
this expression well-defined we analytically continue the chemical potential $\mu$
to imaginary values. For real $x$ we have \cite{R8}
\be
\label{615-46}
\sli^\infty_{n = 1} (-)^n \frac{\cos n x}{n^4}
= \frac{1}{48} \left[ - \frac{7}{15} \pi^4 + 2 \pi^2 x^2 - x^4 \right]  \, .
\ee
Continuing this result back to imaginary values 
$x = i \beta \mu$ we find from eq. (\ref{609-45}) for the pressure $p = - e (\beta)$ 
\be
\label{621-47}
p = \frac{\cN}{24 \pi^2} \left[ \frac{7}{15} \pi^4  T^4 + 2 \pi^2 T^2 \mu^2 + \mu^4 \right] \, ,
\ee
which is the correct result known from the grand canonical ensemble (\ref{388-24}). 
\bi

\no
The equivalence of the expression (\ref{609-45}) for the 
pressure to that of the grand canonical ensemble (\ref{388-24}) can be made explicit by means of the polilogarithm 
\be
\label{627-48}
Li_s (z) = \sli^\infty_{n = 1} \frac{z^n}{n^s} \, ,
\ee
which is defined for arbitrary complex order $s$ and for complex $z$ with $|z| < 1$. By analytic continuation it can be extended to $|z| > 1$. 
The analytically continued form has the integral representation 
\be
\label{633-49}
L i_s (z) = \frac{1}{\Gamma (s)} \il^\infty_0 d t \frac{t^{s - 1}}{e^t / z - 1} \, ,
\ee
by means of which the sum  in eq. (\ref{609-45}) can be expressed as 
\be
\label{638-50}
\sli^\infty_{n = 1} (-)^n \frac{\cos (i n y)}{n^4} = - \frac{1}{12} \il^\infty_0 d t t^3 
\left[ \frac{1}{e^{t - y} + 1} + \frac{1}{e^{t + y} + 1} \right] \, .
\ee
Inserting this relation into eq. (\ref{609-45}) after a change of variables $t = \beta p$ we recover the pressure of the grand canonical ensemble,
eq. (\ref{388-24}), for massless fermions.
\bi

\no
\section{Yang-Mills theory in a quasi-particle approximation}\label{sectIV}
\bi

\no
As a first application of the general method developed in section \ref{section2} to produce new results  we consider Yang-Mills theory at 
finite-temperature in a quasi-particle approximation motivated by the variational calculation in Coulomb gauge \cite{Feuchter:2004mk}, \cite{Epple:2006hv}. 
In this approximation the pseudo-energy density on $\RR^2 \times S^1 (L)$ is 
given by eq. (\ref{401-25}) (or in the regularized form by eq. (\ref{513-34}))
with \cite{Heffner:2012sx} 
\be
\label{780-f15-x1}
\Omega (\vp) = \omega (\vp) - \chi (\vp) \, ,
\ee
where $\omega (\vp)$ is 
the gluon's quasi-particle energy and $\chi (\vp)$ is the ghost loop. For simplicity we will ignore the ghost loop in the 
following. Its influence will be discussed later. In the variational approach in Coulomb gauge 
\cite{Feuchter:2004mk}, \cite{Epple:2006hv}
one finds a gluon quasi-particle energy $\omega (\vp)$, which can be approximated by the so-called Gribov formula
\be
\label{760-x1}
\omega (\vp) = \sqrt{\vp^2 + \frac{m^4}{\vp^2}} \, .
\ee
This formula also nicely fits the lattice data for the gluon propagator
with a Gribov mass of $m = 880 \, MeV$ \cite{Burgio:2008jr}. Unfortunately for this $\omega (\vp)$ 
the pseudo-energy density (\ref{436-30}) cannot be calculated analytically. 
We are interested here in an 
analytic estimate of the pressure for Yang-Mills theory. Therefore, we approximate this expression by the sum
of its infrared and ultraviolet limits
\be
\label{766-x2}
\omega (\vp) = p + \frac{m^2}{p} \, .
\ee
This approximation is applicable in the low and high momentum regime at least but may be too crude in the mid-momentum regime
$p \simeq m$. Therefore we expect that the details of the deconfinement phase transition cannot be adequately reproduced. This refers,
in particular, to the critical temperature, which is sensitive to the details in the mid-momentum regime.
\bi

\no
In a naive attempt one would calculate the pressure and energy density from the pseudo-energy density (\ref{401-25}) or (\ref{436-30}) with
$\Omega (p)$ given by eq. (\ref{766-x2}) resulting in
\be
\label{779-2}
e (\beta) =  
 e_{\alpha = 1} (\beta) + e_{\alpha = - 1}  (\beta)  \, ,
\ee
where we we have defined the finite-temperature momentum integrals
\be
\label{525-11}
e_\alpha (\beta) = \cN \frac{1}{2} \il_\beta \dk {}^3 p \omega_\alpha (p)
\ee
of powers of the 3-momentum
\be
\label{530-12}
\omega_\alpha (p) = m^{1- \alpha} p^\alpha \, , \quad \quad p = \sqrt{\vp^2_\perp + p^2_n} \, .
\ee
Using
the proper-time representation (\ref{482-9})
\be
\label{554-11}
p^\alpha = \frac{1}{\Gamma \lk - \frac{\alpha}{2} \right)} \il^\infty_0 d \tau \tau^{- 1 - \frac{\alpha}{2}} 
e^{- \tau p^2}
\ee
and
the Poisson resummation (\ref{495-10}) where one has to skip the $k = 0$ term 
one finds after carrying out the momentum integrals  $\int \dk {}^2 p_\perp$ and $\int d p_3$ the 
following result 
\be
\label{537-13}
e_\alpha (\beta) =  \cN \frac{m^{1 - \alpha}}{(4 \pi)^{3/2}} \cdot \frac{\Gamma \lk \frac{\alpha}{2} + 
\frac{3}{2} \right)}{\Gamma \lk - \frac{\alpha}{2} \right)} \lk \frac{2}{L} \right)^{3 + \alpha}
\zeta_{n_F} (\alpha + 3) \, ,
\ee
where
\be
\label{543-15}
\zeta_{n_F} (x) = \sli^\infty_{n = 1} \frac{(-1)^{n_F n}}{n^x} \, .
\ee
For bosons $(n_F = 0)$ this quantity is the 
 Riemann $\zeta$-function (\ref{632-13}), $
 \zeta_{n_F = 0} (x) = \zeta (x)$,
while for fermions $(n_F = 1)$ we have
 \be
 \label{581-11-2}
 \zeta_{n_F = 1} (z) = \sli^\infty_{n = 1} \frac{(-1)^n}{n^z} = - \lk 1 - 2^{1 - z} \right) \zeta (z) 
 \, , \quad \quad Re \, z > 0 \, .
\ee
We are interested here in the gauge bosons with dispersion relation (\ref{766-x2}), for which the pressure $p  = - e (\beta)$ 
is given by eqs. (\ref{779-2}), (\ref{525-11}) with degeneracy factor $\cN = 2 (N^2_C - 1)$ for the gauge group $SU (N_C)$.
For bosons we find from eq. (\ref{537-13})
\begin{align}
 \label{784-3}
 e_{\alpha = 1} (\beta) &= -\cN \frac{\pi^2}{90} T^4 \, , \quad \quad
 e_{\alpha = - 1} (\beta) = \cN \frac{m^2}{12} T^2 \, .
\end{align}
The resulting pressure (\ref{779-2}) 
\be
\label{836-17}
p = - e (\beta) = p_{SB} \lk  1 - \frac{45}{6} \frac{m^2}{T^2} \right)
\ee
approaches the correct Stefan-Boltzmann limit 
\be
\label{841-17-1}
p_{SB} = \cN \frac{\pi^2}{90} T^4
\ee
for 
high temperatures but is negative for small temperatures. A negative pressure usually indicates that the underlying 
phase is unstable. Indeed, the quasi-particle vacuum of Yang-Mills theory is unstable against the formation of a constant background
field. In fact, a constant background field aligned along the compactified dimension lowers the (pseudo-) energy density\footnote{A 
constant background field along the uncompactified dimension has no physical effect.}. This is seen by calculating the 
effective potential of such a background field \cite{R4A}. In the Hamiltonian approach the effective potential of a (constant) background
field $\va$ is given by the energy density (here on $\RR^2 \times S^{1} (L)$ by the pseudo-energy density), calculated in 
the presence of the constraint that the expectation value of the dynamical field $\vA$ equals the background field \cite{R25}
\be
\label{797-4}
\langle \vA \rangle = \va \, .
\ee
Without loss of generality the background gluon field can be chosen in the Cartan subalgebra. Up to two loop order the effective potential of 
the background field is then obtained from the (pseudo-)energy density (\ref{401-25}) by shifting the momentum \cite{R4A}
\be
\label{803-4-1}
\vp \to \vp_\sigma = \vp - \vec{\sigma} \va = \vp_\perp + \ve_3 (p_n - \vec{\sigma} \va) \, ,
\ee
where $\vec{\sigma}$ denotes a root vector of the gauge group, and summing over all roots
\be
\label{808-4-3}
e (a, L) =  \sli_{\vec{\sigma}} \il_L \ddbar^3 p \Omega (p_\sigma) \, .
\ee
Here the degeneracy factor $\cN = 2 (N^2_C - 1)$ is already included. The factor $2$  stemming from the two polarization degrees 
of freedom cancels the factor $1/2$ in equation (\ref{401-25}) while the color degeracy factor 
$N^2_C - 1$ is accounted for by the summation over the root vectors $\vec{\sigma}$. 
\bi

\no
To find the minimum of the effective potential $e (a, \beta)$ it is convenient to subtract the pseudoenergy density at vanishing 
background field $e (a = 0, \beta) = e (\beta)$. The difference 
\be
\label{814-f17-x1}
e_p (a, \beta) := e (a, \beta) - e (a = 0, \beta) \, 
\ee
is ultraviolet finite and can be interpreted as 
 the effective potential of the 
Polyakov loop  \cite{R4A}.
To obtain the Polyakov loop potential in the quasi-particle approximation for the dispersion relation (\ref{766-x2}) it 
is conventient to calculate first the extension of $e_\alpha (\beta)$ (\ref{525-11}) in the presence of the background 
field
\be
\label{825-x2}
e_\alpha (a, \beta) = \sli_{\vec{\sigma}} \il_L \ddbar^3 p \omega_\alpha (\vp_\sigma) \, ,
\ee
where $\omega_\alpha (p)$ is defined in eq. (\ref{530-12}). 
The calculations are done in the same way as in absence of the background field except that after doing the Poisson resummation 
the integration variable $p_3$ is shifted: $p_3 - \vec{\sigma} \cdot \va \to p_3$.  One finds
\be
\label{831-x3}
e_\alpha (a, \beta) = 2  \frac{m^{1 - \alpha}}{(4 \pi)^{3/2}} \frac{\Gamma \lk \frac{3 + \alpha}{2} \right)}{\Gamma \lk - \frac{\alpha}{2} \right)}
\lk \frac{2}{\beta} \right)^{\alpha + 3} h_\alpha (a, \beta) \, ,
\ee
where
\be
\label{837-x4}
h_\alpha (a, \beta) = \sli_{\vec{\sigma}} \sli^\infty_{n = 1} \frac{\cos (n \vec{\sigma} \cdot \va \beta)}{n^{\alpha + 3}} \, .
\ee
Since 
\be
\label{842-f17b}
h_\alpha (a = 0, \beta) = (N^2_C - 1) \zeta (\alpha + 3)
\ee
the expression $e_\alpha (a, \beta)$ (\ref{831-x3}) reduces for $a = 0$ indeed to $e_\alpha (\beta)$ (\ref{537-13}). 
For the dispersion relation (\ref{766-x2}) we find for the Polyakov loop potential
\be
\label{848-71b-1}
e_p (a, \beta) = e_{\alpha = 1} (a, \beta) + e_{\alpha = - 1} (a, \beta) - \lk e_{\alpha = 1} (0, \beta) + e_{\alpha = - 1} (0, \beta) 
\right) \, .
\ee
The various pieces can be evaluated analytically. For the gauge group $SU(N_C = 2)$ the roots are one-dimensional
\be
\label{853-f17}
\vec{\sigma} \va = \sigma a \, , \quad \quad \sigma = 0 \,,  \, \pm 1 \, ,
\ee
so that
\be
\label{858-f17-1}
h_\alpha (a, \beta) = 2 \sli^\infty_{n = 1} \frac{\cos (n a \beta)}{n^{\alpha + 3}} + \zeta (\alpha + 3) \, .
\ee
Using \cite{R8}
\be
\label{863-17-2}
\sli^\infty_{k = 1} \frac{\cos (k x)}{k^2} = \frac{\pi^2}{6} - \frac{\pi x}{2} + \frac{x^4}{4} 
\ee
and $\zeta (2) = \pi^2 / 6$ one finds for $\alpha = - 1$
\be
\label{868-17-2}
e_{\alpha = - 1} (a, \beta) =  (N^2_C - 1) \frac{m^2}{6} T^2 \lk 1 - \frac{a}{\pi T} \right)^2 \, .
\ee
For $\alpha = 1$ we use \cite{R8}
\be
\label{873-17-3}
\sli^\infty_{k = 1} \frac{\cos (kx)}{k^4} = \frac{\pi^4}{90} - \frac{\pi^2 x^2}{12} + \frac{\pi x^3}{12} - \frac{x^4}{48}
\ee
and $\zeta (4) = \pi^4/90$ to obtain 
\be
\label{878-f17-4}
e_{\alpha =  1} (a, \beta) = -  (N^2_C - 1) \frac{\pi^2 T^4}{45} \left[ 1 - 20 \lk \frac{a}{2 \pi T} \right)^2 \lk 1 - \frac{a}{2 \pi T} \right)^2 \right] \, .
\ee
With the explicit expressions for $e_{\alpha = \pm 1} (a, \beta)$ at hand we find for the effective potential of the 
Polyakov loop (\ref{848-71b-1})
\be
\label{884-17-x1}
e (a, \beta) = (N^2_C - 1) \frac{4}{9} \pi^2 T^4 f (x) \, ,
\ee
where
\be
\label{889-17-x2}
f    (x) = x^2 (x - 1)^2 + c x (x - 1)
\ee
with the dimensionless variables 
\be
\label{894-17-x3}
x = \frac{a \beta}{2 \pi} \, , \quad \quad c = 3 \frac{m^2 \beta^2}{2 \pi^2} \, .
\ee
Note that the potential (\ref{884-17-x1}) is invariant under the center 
transformation $x \to 1 - x$.
\bi

\no
For $c > \frac{1}{2}$ the function $f (x)$ (\ref{884-17-x1}) has a single real 
extremum, i.e. a minimum at $x = \frac{1}{2}$. This minimum turns into a degenerate cubic root at $c = \frac{1}{2}$ and 
eventually
for $c < \frac{1}{2}$ dissolves into a maximum at $x = \frac{1}{2}$ and two degenerate 
minima at 
\be
\label{902-17-x4}
x_{1/2} = \frac{1}{2} \left[ 1 \pm \sqrt{1 - 2c} \right] \, ,
\ee
which are related by a center transformation
$x \to 1 - x$, i.e. $x_2 = 1 - x_1$. 
\bi

\no
For given temperature $T = 1/\beta$ (i.e. for given value of $c$ (\ref{894-17-x3})) the vacuum background field configuration $\bar{a}$ 
is given by the minimum of the Polyakov  loop potential (\ref{884-17-x1}).
For $c > \frac{1}{2}$ the minimum occurs at the center symmetric point $x = \frac{1}{2}$, i.e. $\bar{a} = \frac{\pi}{\beta}$ 
corresponding to the confined phase. 
At $c = \frac{1}{2}$ where this minimum turns into a maximum the deconfinement phase transition occurs. From eq. (\ref{894-17-x3}) we 
find for the critical temperature
\be
\label{910-17-5}
T_c = \sqrt{3} m / \pi \, .
\ee
For a Gribov mass of $m \simeq 880 \, MeV$ \cite{Burgio:2008jr}, which fits the lattice data for the gluon propagator one finds $T_c \simeq 485 \, MeV$,
which is by far too high compared to the lattice result of $T_c \simeq 300 MeV$. As shown in ref. \cite{R4A} the high value of $T_c$ results from the 
neglect of the ghost loop in the (pseudo-)energy density. 
Inclusion of the ghost loop lowers the critical temperature to realistic values \cite{R4A}. 
\bi

\begin{center}
  \begin{figure}
 \includegraphics[width=.6\textwidth]{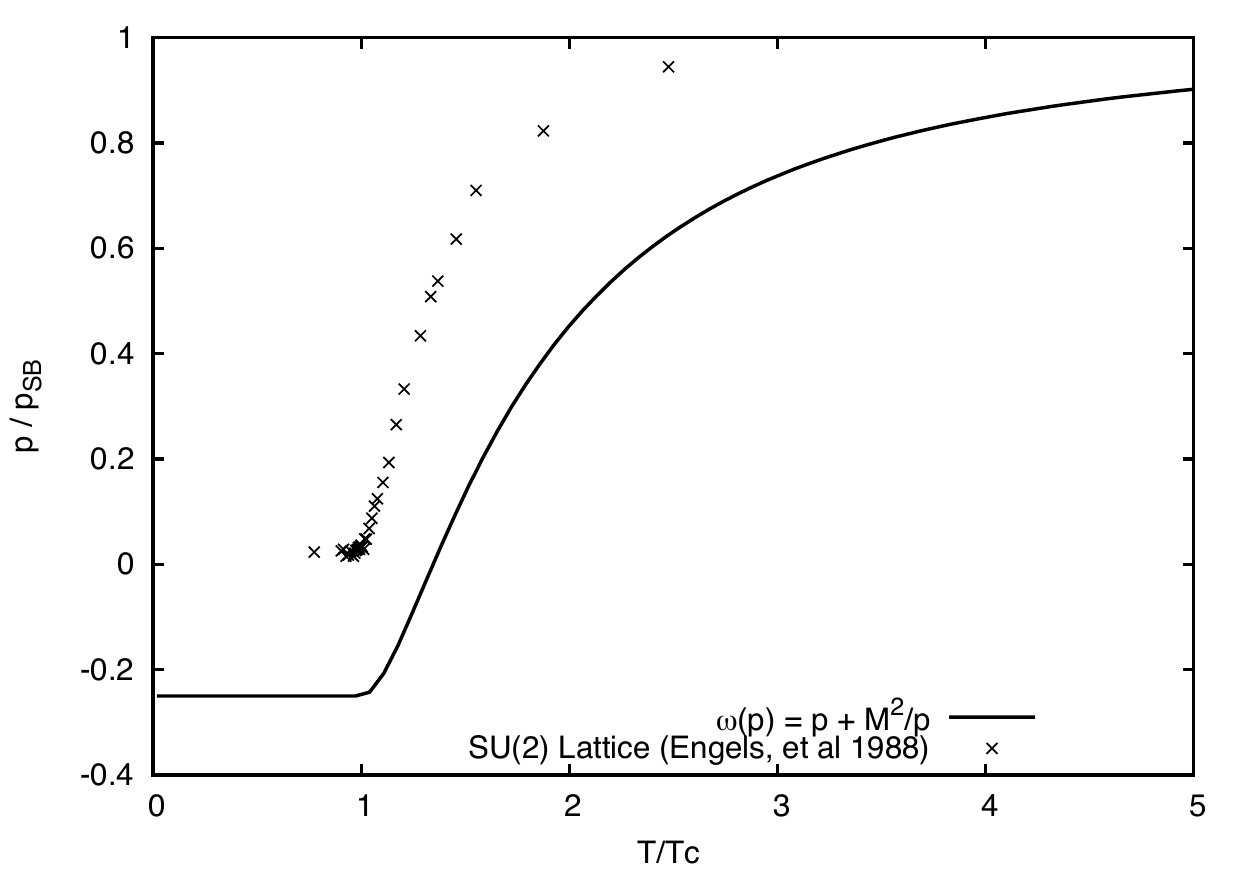}
 \caption{The pressure $p$ of Yang-Mills theory calculated by the method of sect.~\ref{section2} for a gluon dispersion relation 
 $\omega (p) = p + M^2/p$. Shown is the ratio $p/p_{SB}$ where $p_{SB}$ denotes the Stefan-Boltzman  limit (\ref{931-17-8}). 
 The crosses show the lattice data from ref. \cite{Engels:1988ph}. }\label{pressure}
\end{figure}
\end{center}

\no
By means of the critical temperature $T_c$ (\ref{910-17-5}) we can express the quantity $c$ (\ref{894-17-x3}) as
\be
\label{946-19-1}
2 c = \lk \frac{T}{T_c} \right)^2 \, ,
\ee
so that the two degenerate minima (\ref{902-17-x4}) of the   effective potential $e_p (a, \beta)$ 
at $T > T_c$ read
\be
\label{946-19}
x_{1/2} = \frac{1}{2} \left[ 1 \pm \sqrt{1 - \lk \frac{T_c}{T} \right)^2} \right] \, .
\ee
The pressure of Yang-Mills theory is given by
\be
\label{906-17-4}
p = - e (\bar{a}, \beta) = - e_{\alpha = - 1} (\bar{a}, \beta) - e_{\alpha = 1} (\bar{a}, \beta) \, ,
\ee
where $\bar{a}$ is the position of the minimum of the Polyakov loop potential, which (in the dimensionless variable $x$ 
(\ref{894-17-x3}) is given by
\be
\label{911-17-5}
\bar{x} = \left\{ \begin{array}{ccc}
                   \frac{1}{2} & , & T \leq T_c \\
                   x_{1/2} & , & T \geq T_c
                  \end{array}
                  \right. \, .
                  \ee
In the  confined phase we have $\bar{x} = 1/2$ or $\bar{a} = \pi/\beta$ and 
\be
\label{920-17-6}
e_{\alpha = - 1} \lk \bar{a} = \frac{\pi}{\beta}, \beta \right) = 0 \, , \quad \quad e_{\alpha = 1} \lk 
\bar{a} = \frac{\pi}{\beta}, \beta \right) =  (N^2_C- 1) \frac{\pi^2}{180} T^4 \, , 
\ee
so that we find for the pressure
\be
\label{926-17-7}
p (t) = - \frac{1}{4} p_{SB} (t) \, , 
\ee
where 
\be
\label{931-17-8}
p_{SB} (t)  = - e_{\alpha = 1} (a = 0, \beta) = (N^2_C - 1) \frac{M^4}{5 \pi^2} t^4 
\ee
is the Stefan-Boltzmann limit (\ref{841-17-1}) of the pressure and we have introduced the dimensionless temperature $t = T /T_c$. In the 
deconfined phase we obtain with
\begin{align}
 \label{1011-f20}
 e_{\alpha = - 1} (x_{1/2}, \beta) & = (N^2_C - 1) \frac{m^4}{2 \pi^2} t^2 \lk 1 - \frac{1}{t^2} \right) \, , \nonumber\\
 e_{\alpha = 1} (x_{1/2}, \beta) &= -  (N^2_c - 1) \frac{m^4}{\pi^2} t^4 \lk \frac{1}{5} - \frac{1}{4 t^2} \right)
\end{align}
for the pressure
\be
\label{952-17-5}
p (t) = p_{SB} (t) \left[ 1 - \frac{15}{4} \frac{1}{t^2} + \frac{5}{2} \frac{1}{t^4} \right] \, .
\ee
For $T \to \infty$ the pressure approaches the Stefan-Boltzmann limit. The pressure  obtained above is shown in fig. \ref{pressure} as 
function of the temperature. For sake of comparison we also show the pressure measured on the lattice \cite{Engels:1988ph}, where 
the scale was adjusted to match the critical temperature $T_c$. Above $T_c$ the tendency of the lattice data is 
roughly reproduced. Given the crudness of our approximation we cannot expect a good agreement with the lattice data. 
In the confined phase we still get a negative pressure
\be
\label{1053-17-2}
p / p_{SB} = - 1/4 \, , 
\ee
which, however, is much more benign than the result of the naive calculation (\ref{836-17}), for which
\be
\label{1058-20-1}
p / p_{SB} \sim - 1/T^2 \, , \quad \quad T \to 0 \, .
\ee
The negative pressure obtained in eq.~(\ref{1053-17-2})  for $T < T_c$ 
is not due to an instability of the confined phase but rather a
consequence of the violation of $O (4)$ invariance by the dispersion relation (\ref{766-x2}).\footnote{Please recall that the 
present approach to finite-temperature quantum field theory by compactifying a spatial dimension assumes $O (4)$ invariance
in Euclidean space, 
see sect. \ref{section2}. In the cases treated in sect. \ref{sectIII} the $O (4)$ invariance was strictly preserved and the exact 
results were obtained.} This is seen by calculating numerically the pressure for the quasi-particle energy (\ref{766-x2}) 
from the grand canonical ensemble (\ref{486-f9-1}), which yields a positive definite result \cite{R23}. 
Since both approaches, the grand canonical ensemble and the compactification of a spatial axis, are equivalent for $O (4)$
invariant theories the  negative pressure obtained above is definitely a consequence of the $O (4)$-violation of the 
dispersion relation (\ref{766-x2}) but not a consequence of a vacuum instability. 


In general, in a Hamiltonian approach approximations may lead to a violation of the $O (4)$-symmetry. In the variational approach 
to Yang-Mills theory in Coulomb gauge \cite{Reinhardt:2004mm,Feuchter:2004mk,Epple:2006hv} it is not difficult to see that the inclusion 
of the ghost reduces the $O(4)$-symmetry breaking. Indeed, in this approach the self-consistent pseudoenergy density is given by 
eq. (\ref{401-25}) with $\Omega (\vp)$
given by eq. (\ref{780-f15-x1}).
The ghost loop $\chi (\vp)$ is infrared  divergent and vanishes in the ultraviolet. If one uses the 
Gribov formula $\omega (\vp)$ (\ref{760-x1}) and ignores the so-called 
Coulomb term, the variational gap equation yields for the ghost loop \cite{R4A}
\be
\label{1180-2}
\chi (\vp) = m^2 / |\vp| \, .
\ee
Using the approximation eq. (\ref{766-x2}) to the Gribov formula and eq.  (\ref{1180-2})
we obtain from (\ref{780-f15-x1}) $\Omega (p) = |\vp|$,  which is an 
$O(4)$-invariant dispersion relation. Of course, in a realistic calculation using the 
numerical variational solutions the inclusion of the ghost will not completely restore the $O(4)$-symmetry 
but will definitely reduce the symmetry breaking \cite{R23}.

\section{Summary and Conclusions}\label{sectV}
\bi

\no
I have presented an alternative approach to finite-temperature quantum field theory within the Hamiltonian formulation where the 
temperature is introduced by compactifying a spatial dimension. Compared to the usual grand canonical ensemble this approach is
advantageous in the sense that it does not require the introduction of a statistical density operator. Instead the whole temperature
behaviour is encoded in the vacuum state on the spatial manifold $\RR^2 \times S^1$. This is beneficial for non-perturbative continuum 
studies like variational approaches, which usually concentrate on the description of the vacuum while excited states are not directly accessible.
I have illustrated this approach for free bosons and fermions where it reproduces the correct result of the grand canonical ensemble. 
Furthermore, the pressure of Yang-Mills theory was calculated in a quasi-particle approximation using quasi-gluon energies, 
which were motivated by the results of a variational approach in Coulomb gauge and also by lattice data for gluon propagator in 
Coulomb gauge. In order to carry out the calculations analytically we used a simple parametrization of the 
quasi-gluon energy, which unfortunately violates $O (4)$-invariance, in particular at small momenta. As a consequence we obtained a small negative 
pressure in the confined phase. The $O (4)$ symmetry breaking of the assumed gluon energy (\ref{766-x2}) decreases with increasing momenta
and the obtained pressure reaches
the correct Stefan-Boltzmann limit at high temperature.
\bi

\no
Let us also mention that the present
approach was applied in ref. \cite{Heffner:2015zna} to study finite-temperature Yang-Mills theory 
in a variational approach and in ref. \cite{R4A} to calculate the effective potential of the 
Polyakov loop (in pure Yang-Mills theory) at finite-temperature. In the future I plan to apply this approach also to the quark sector of QCD.
\bi

\no
\section*{Acknowledgement} 
The author thanks D. Campagnari, J. Heffner and M. Quandt for useful discussions. Furthermore, he thanks J. Heffner for providing the figures. This work 
was supported by Deutsche Forschungsgemeinschaft under contract DFG-Re856/9-2.
\bi

\no

\end{document}